\documentclass[8.5pt,twoside,twocolumn]{article}
\usepackage[super,sort&compress,comma]{natbib} 
\usepackage{mhchem}
\usepackage{times,mathptmx}
\usepackage{sectsty}
\usepackage{balance} 

\usepackage{graphicx} 
\usepackage{lastpage}
\usepackage[format=plain,justification=raggedright,singlelinecheck=false,font=small,labelfont=bf,labelsep=space]{caption} 
\usepackage{fancyhdr}
\usepackage[utf8]{inputenc}
\usepackage[amssymb]{SIunits}
\usepackage{amssymb,amsmath}
\usepackage{notes2bib}
\usepackage{booktabs}

\usepackage{xspace}
\newcommand{\SLcam}{SLcam\textsuperscript{\textregistered}\xspace}

\begin{document}

\thispagestyle{plain}
\fancypagestyle{plain}{
\renewcommand{\headrulewidth}{1pt}
}
\renewcommand{\thefootnote}{\fnsymbol{footnote}}
\renewcommand\footnoterule{\vspace*{1pt}%
\hrule width 3.4in height 0.4pt \vspace*{5pt}} 
\setcounter{secnumdepth}{5}

\makeatletter 
\def\subsubsection{\@startsection{subsubsection}{3}{10pt}{-1.25ex plus -1ex minus -.1ex}{0ex plus 0ex}{\normalsize\bf}} 
\def\paragraph{\@startsection{paragraph}{4}{10pt}{-1.25ex plus -1ex minus -.1ex}{0ex plus 0ex}{\normalsize\textit}} 
\renewcommand\@biblabel[1]{#1}            
\renewcommand\@makefntext[1]%
{\noindent\makebox[0pt][r]{\@thefnmark\,}#1}
\makeatother 
\renewcommand{\figurename}{\small{Fig.}~}
\sectionfont{\large}
\subsectionfont{\normalsize} 

\fancyfoot{}
\fancyfoot[RO]{\footnotesize{\sffamily{1--\pageref{LastPage}~\textbar  \hspace{2pt}\thepage}}}
\fancyfoot[LE]{\footnotesize{\sffamily{\thepage~\textbar~
 1--\pageref{LastPage}}}}
\fancyhead{}
\renewcommand{\headrulewidth}{1pt} 
\renewcommand{\footrulewidth}{1pt}
\setlength{\arrayrulewidth}{1pt}
\setlength{\columnsep}{6.5mm}
\setlength\bibsep{1pt}

\twocolumn[
  \begin{@twocolumnfalse}
\noindent\LARGE{\textbf{%
Sub-pixel resolution with color X-ray camera \SLcam
}}
\vspace{0.6cm}

\noindent\large{\textbf{%
Stanisław H. Nowak,$^{\ast}$\textit{$^{a}$} 
Aniouar  Bjeoumikhov,\textit{$^{ab}$}
Johannes von Borany,\textit{$^{c}$} 
Josef Buchriegler,\textit{$^{c}$} 
Frans Munnik,\textit{$^{c}$} 
Marko Petric,\textit{$^{d}$} 
Martin Radtke,\textit{$^{e}$}  
Axel D. Renno,\textit{$^{f}$} 
Uwe Reinholz,\textit{$^{e}$} 
Oliver Scharf,\textit{$^{a}$}  
and 
Reiner Wedell\textit{$^{g}$} 
}}\vspace{0.5cm}

%

\noindent \normalsize{%
The color X-ray camera \SLcam is a full-field, single photon detector providing scanning free, energy and spatially resolved X-ray imaging. 
Spatial resolution is achieved with the use of polycapillary optics 
guiding
X-ray photons from small regions on a sample  to distinct energy dispersive pixels on a charged-coupled device detector. 
Applying  sub-pixel resolution, signals from individual capillary channels can be distinguished.
Accordingly the \SLcam spatial resolution  can be released from  pixel size being confined rather to a diameter of individual polycapillary channels.
In this work a new approach to sub-pixel resolution algorithm comprising   photon events also from the pixel centers is proposed. 
The details of the employed numerical method and  several sub-pixel resolution examples are presented and discussed.
}
\vspace{0.5cm}
 \end{@twocolumnfalse}
  ]



\footnotetext{\textit{$^{a}$~IfG -- Institute for Scientific Instruments GmbH, Berlin, Germany.
E-mail: nowak@ifg-adlershof.de
}}
\footnotetext{\textit{$^{b}$~Institute for Computer Science and Problems of Regional Management,
 Kabardino-Balkaria, Russia. }}
\footnotetext{\textit{$^{c}$~Helmholtz-Zentrum Dresden-Rossendorf, Dresden, Germany. }}
\footnotetext{\textit{$^{d}$~J. Stefan Institute, Ljubljana, Slovenia.}}
\footnotetext{\textit{$^{e}$~BAM Federal Institute for Material Research and Testing, Berlin, Germany.}}
\footnotetext{\textit{$^{f}$~Helmholtz-Zentrum Dresden-Rossendorf, Helmholtz Institute Freiberg for Resource Technology,
Freiberg, Germany.}}
\footnotetext{\textit{$^{g}$~IAP Institute for Applied Photonics e.V., Berlin, Germany.}}

\section{Introduction}

\SLcam  is a high quantum efficiency and throughput color X-ray camera system\cite{Kuehn2011,Ordavo2011250,Scharf2011} 
 designed for divergent X-ray radiation.
It  allows detection of single photons with both energy and spatial resolution.
\SLcam combines a pn-junction Charged-Coupled Device (pnCCD) \cite{Strueder2001} with polycapillary
optics.\cite{Bjeoumikhov2003}
The latter can be regarded as a bunch of independent X-ray channels guiding X-ray photons  from  small regions on a sample to corresponding pixels on pnCCD similarly to the way as fiber
optics guide light.


%

\SLcam employs pnCCD with \unit{48}{\mu m} pixel size. 
Keeping the pixel size fixed spatial resolution of \SLcam can be seriously improved by the use of
conically shaped magnifying optics. \cite{Scharf2011,Ordavo2011250}
This type of optics is  available with  magnification factors up to 10:1,   allowing representation of
4.8$\times$\unit{4.8}{\mu m^2} area in a single pixel. 
Theoretically, according to Nyquist-Shannon sampling theorem, \cite{Shannon1949} such a system can correctly resolve details down to \unit{\sim\!9}{\mu m}.

Currently used \SLcam optics are optimized for the pnCCD pixel dimension.
The capillary exit diameter is adapted in such a way that the spot size
from an individual channel on the detector is approximately equal to the pixel size.
However, current technology allows fabrication of polycapillaries with single channel diameter in the range of a micron or even below  giving room for further improvement  of  resolution. 
Currently pixel size is the limiting factor for  spatial resolution of  \SLcam.   

With the use of a sub-pixel algorithm the dominant role of  pixel size can be released. 
This algorithm  divides the signal assigned to each physical pixel over a number of virtual sub-pixels. 
With such an approach further downscaling of polycapillary channels can be practically used for \SLcam lateral resolution improvement.

 Images with sub-pixel resolution are achievable
  due to
 specific physics of X-ray photon with Charged-Coupled Device (CCD)  interaction that guides to creation of a so-called electron cloud.
 A non zero area of the cloud leads to charge deposition in  pixels nearest to the photon hit. 
 With a correct reconstruction of the footprint of a single photon event the photon hit position can be estimated with a much higher precision than the pixel size.

The sub-pixel resolution algorithm was first applied to Advanced CCD
Imaging Spectrometer (ACIS) installed at Chandra X-Ray
Observatory (CXO). \cite{Tsunemi1997,Pivovaroff1998,Tsunemi1999,Yoshita1999,Hiraga2001,Li2003}
Some years later similar technique was adapted to pnCCD. \cite{Kimmel2006,Abboud2013} 
Unfortunately, the proposed routines are neglecting a majority of events analyzing only  so-called "corner events", {\it i.e.}, the photon hits that reach the CCD in  proximity to a pixel corner.
As a result the sub-pixel image exhibit a grid pattern  with the intensity drops in pixel centers.\cite{Scharf2011}
We propose a modified approach taking all the photon hits and distributing them properly over the sub-pixel pattern.
 
\section{Method}

\subsection{Electron cloud}

An X-ray photon absorbed in silicon generates a number of electron-hole pairs amounting to  $E/W_{Si}$ -- the photon energy $E$ divided by  the  formation energy of a single
electron-hole pair $W_{Si} = \unit{3.6}{eV}$. \cite{Scholze1998}
%
In a fully depleted  layer the carriers are separated in the vertical electric field and diffuse laterally producing two charge clouds  with opposite signs.
The holes are collected at the large area cathode. 
The electrons are transfered to the pixelated anode  where they are split over individual pixels. During the readout these charge packets  are sequentially transferred to charge amplifiers and counted.

%


The carrier density of a single photon electron cloud is frequently approximated  by a two dimensional Gaussian distribution \cite{Janesick2001,Hopkinson1987,Hiraga2001,Miyata2003}:
\begin{equation}
 S(x,y ) = {Q\over 2\pi\sigma_x\sigma_y} \exp\left( - {(x-x_0)^2\over 2
 \sigma_x^2}-{(y-y_0)^2\over 2
 \sigma_y^2 } \right),
\end{equation}
where $Q$ is the total charge produced, $\sigma_x$ and $\sigma_y$ are the $x$ and $y$ widths of the cloud, and $(x_0,y_0)$ is the point of a photon hit. 
For a pixel with coordinates $(i,j)$ the resulting charge intensity $I^{i}_{j}$ can be calculated as an integral over a square corresponding to a pixel area:
\begin{equation}
\label{eq:intesity}
\begin{aligned}
\begin{aligned}
I^{i}_{j} &=\   \int_{x_i}^{x_{i+1}}\!\!\!\int_{y_{j}}^{y_{j+1}} S(x,y)\ dx\,dy\ =\\
 &=\  {Q\over 4} \left(\operatorname{erf}\ {x_{i+1}-x_0\over \sqrt{2}\,\sigma_x} -
\operatorname{erf}\ {x_{i}-x_0\over \sqrt{2}\,\sigma_x} \right)
\end{aligned}
\\
\cdot\left(\operatorname{erf}\
{y_{j+1}-y_0\over \sqrt{2}\,\sigma_y} - \operatorname{erf}\ {y_{j}-y_0\over
\sqrt{2}\,\sigma_y} \right).
\end{aligned}
\end{equation}
Note that  $I^{i}_{j}$  is a product of two separable functions depending only on one variable $x$ or $y$.

Though this is a simplified model neglecting the energy dependence, charge quantization,  {\it etc.}, it shows the fundamental fact making the sub-pixel resolution algorithm possible: on a pixelated plane a single photon electron cloud is normally split over several pixels. 
For every photon hit, in addition to pixel position,  extra information in a form of pixel intensity distribution is gathered.
This additional information can be used to determined the center of impact of the photon 
 with sub-pixel accuracy.

\subsection{Intensity ratios}
A good measure of a single photon electron cloud distribution are the intensity ratios defined as follows:
\begin{equation}
\label{eq:intensity_datio_1}
d(I_1,I_2) = \frac{I_2 - I_1}{I_2+I_1},
\end{equation}
where $I_1$ and $I_2$ correspond to  charge gathered in two different areas of a CCD. 
Possible values of $d(I_1,I_2)$ ranges  from $-1$ to $1$; 
$0$ is reached when $I_1=I_2$.

It can be shown that for any electron cloud distribution for which x and y components can be separated ({\it e.g.}, Gaussian distribution)  the pixel intensity ratio in $x$ direction is independent from $y$ coordinate and {\it vice versa}:
\begin{eqnarray}
 d(I^{i-1}_{j},\ I^{i}_{j}) \;\;=\;\; d(I^{i-1}_{k},\ I^{i}_{k}), &\operatorname{\ and} \\
 d(I^{i}_{j-1},\ I^{i}_{j}) \;\;=\;\; d(I^{l}_{j-1},\ I^{l}_{j}).
\end{eqnarray}
Note also that, provided that 
for each photon hit
the shape of the electron cloud is uniform, the intensity ratio should change monotonically with $x$ and $y$ coordinates of the hit position. 



A very elegant way to estimate the single photon electron cloud distribution was presented in Ref~\citenum{Lawrence2011}. 
The method relies on analysis of a
histogram of
measured intensity ratios  
and assumes only a uniform distribution of  photon hits over the CCD plane and a Gaussian shape of the charge cloud. 
The method was used to reveal the size of the cloud.
We will use similar methodology  
to calculate sub-pixel coordinates. 

In order to calculate the intensity ratio histogram  
2$\times$2 pixel boxes centered at a pixel corner $(x_i,y_j)$ nearest to the point of a  photon hit $(x_0,y_0)$
are analyzed;
intensity ratios in, respectively, $x$ and $y$ directions
are calculated as follows:
\begin{eqnarray}
\label{eq:d_x}
d_x&=&  d(I^{i-1}_{j-1}+I^{i-1}_{j},\ I^{i}_{j-1}+I^{i}_{j}) \\
\label{eq:d_y}
d_y& =&  d(I^{i-1}_{j-1}+ I^{i}_{j-1},\ I^{i-1}_{j}+ I^{i}_{j}).
\end{eqnarray}
The intensity ratios in $x$ or $y$ direction computed for all the single photon hits are combined in one histogram.

The histogram normalized to occupy a unitary area
can be regarded as an estimation of  the probability density function.
Accordingly, a normalized intensity ratios histogram  rates the probability of a single photon hit to create a given intensity ratio value.

\begin{figure}[!b]
\centering
\includegraphics[width=\linewidth]{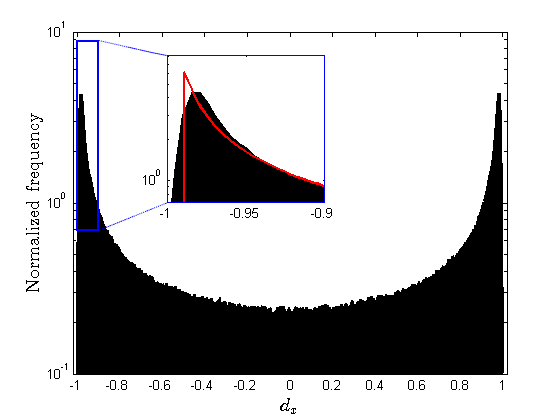}
\caption{An example intensity ratio histogram of Au L photons  generated from a real measurement (Au bar pattern) with formula~\eqref{eq:d_x}. 
The histogram is normalized to occupy a unitary area
and can be identified with the probability density function.
In the inset the histogram is compared to  Gaussian simulation (red line). 
}
\label{fig:Difference_ratio_histogram}
\end{figure}

In Fig.~\ref{fig:Difference_ratio_histogram} an example of normalized intensity ratio histogram  is plotted. 
The figure shows data computed for Au~L photons from the measurement of Au bar pattern presented further in  Section~\ref{sec:Experimental}.
As can be seen most of the photon hits create electron clouds with the intensity ratio in proximity to $-1$ or $1$. 
Intensity ratios close to $0$ are the least probable.
Discussed Gaussian model fits well to most values of   the histogram;
though for $|d_x| \approx 1$ (and $|d_y| \approx 1$) the simulation fails showing deficiency of the model.

\subsection{Sub-pixel coordinates}
Our method consist in converting the $d_x$ and $d_y$ intensity ratios of a given photon hit to $x_0$ and $y_0$ coordinates.
In our approach 
we do not model
the shape of the single photon electron cloud;
we assume only
separation of $x$ and $y$ components of the charge distribution.
We also assume a  
uniform distribution of photon hits over the CCD plane.
Accordingly the relation between the intensity ratio and single photon hit position is not simulated, but measured.

Lets set $(x_i,y_j)$ -- the pixel corner  nearest to the point of a  photon hit --
 to the origin, and the number of photon events to $N$.
The possible values for $x_0$ inside a 2$\times$2 pixel box span within $\left(-{p\over 2},{p\over 2}\right)$,  where $p$ is the pixel dimension. 
Due to  uniform distribution of  photon hit positions over the CCD plane 
the probability of finding a photon hit with  $x$ coordinate  below $x_0$ has a linear form:
\begin{equation}
P_x(x_0)  
		 = {n( x\!:\  x < x_0) \over N}
		 = {x_0\over p} + {1\over 2};
\end{equation}
here $n()$ stands for the number of elements of a set.
%
%
%
We can also compute 
cumulative probability functions   of $d_x$ and $d_y$:
\begin{eqnarray}
\label{eq:intensity_datio_2}
P_{d_x}(d_x) &=& {n\left(\tilde d_x\!:\ \tilde d_x < d_x\right)\over N},\\
P_{d_y}(d_y) &=& {n\left(\tilde d_y\!:\ \tilde d_y < d_y\right) \over N }.
\end{eqnarray}
$P_{d_x}(d_x)$ and $P_{d_x}(d_x)$ are equivalent to normalized cumulative histograms of the measured intensity ratios.

If we  note that   $d_x$ is  monotonically increasing with $x_0$
then we get:
\begin{equation}
n\left(\tilde d_x\!:\ \tilde d_x < d_x(x_0)\right)= n\left( x\!:\  x < x_0\right).
\end{equation}
This directly  shows that   $P_{d_x}$ can be identified with $P_x$:
\begin{equation}
P_{d_x}(d_x(x_0)) = P_x(x_0).
\end{equation}
Finally  an explicit relation for $x_0$ can be given:
\begin{equation}
\label{eq:x_0}
   x_0 = {p}\left(P_{d_x}(d_x) -{1\over 2}\right).
\end{equation}
An analogical relation is also valid  for $y_0$ and $d_y $.

\begin{figure}
\centering
\includegraphics[width=\linewidth]{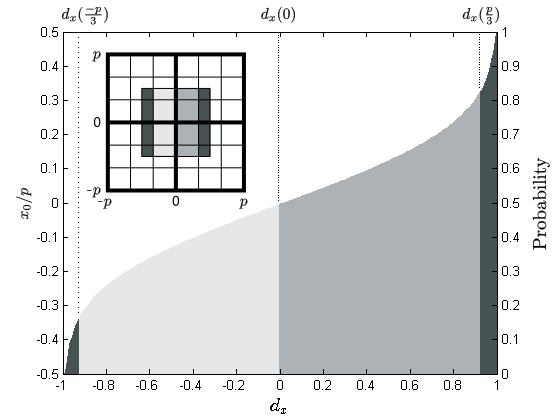}
\caption{
Relative photon hit position $x_0$  calculated as  a function of intensity ratio $d_x$  according to equation~\eqref{eq:x_0}.  
Likewise in Fig.~\ref{fig:Difference_ratio_histogram}
the data for Au L photons from Au bar pattern   are illustrated.
Secondary vertical axis shows cumulative probability distribution $P_{d_x}$.
In the inset a 2$\times$2 pixel box with  3$\times$3 sub-pixel division is shown. 
Different levels of gray mark -- in order from darkest to brightest -- central, left and right sub-pixels.
 }
\label{fig:Difference_ratio_cumsum}
\end{figure}

The relationship between $x_0$, $d_x$ and $P_{d_x}$ is shown in  Fig.~\ref{fig:Difference_ratio_cumsum}.

\subsection{Division to sub-pixels\label{sec:sub_pixel_division}}
A strong point of our approach is that it
requires only a good statistics of single photon hits
ensuring high enough   number of events $N$ to minimize the statistic error of generated $P_{d_x}$ and $P_{d_y}$.
In particular the method is independent from the shape and creation process of the electron cloud
and can  deal with any, also asymmetric, \cite{Tsunemi1999} charge distributions.
The  cloud creation process and the resulting charge distribution is a characteristics of a given CCD and should not change in time. 
Therefore  each measurement could be analyzed with the use of the same, previously obtained 
pixel intensity ratio
probabilities.

In an ideal case of uniformly shaped electron clouds and  a  hypothetical noise-free CCD, equation~\eqref{eq:x_0} should give a strict position of a single photon hit. 
In reality the electron cloud is shaped randomly  and the intensity signal from each pixel is 
given with an error.
These two factors contradict the assumptions on monotonic relation between $d_x$ and $x_0$ and on separation of $x$ and $y$ components.
As a result  the accuracy of sub-pixel coordinates is  limited. 

We can estimate an error on $d(I_1,I_2)$ with the following relation:
\begin{equation}
\delta d(I_1,I_2)= {2 \delta I  \sqrt{I_1\,^2 +I_2\,^2} \over (I_1 + I_2)^2},
\end{equation} 
where  $\delta I$ is an error on a charge gathered in $I_1$ or $I_2$.
Note that $\delta d(I_1,I_2)$ has its maximum for $I_1=0$ or $I_2=0$ which corresponds to $|d(I_1,I_2)|=1$ or $|x_0| = {p\over 2}$.
This is exactly where $P_{d_x}$ exhibits the largest increase and, respectively, where the error on $P_{d_x}$ is the biggest. 

As a result the photon hit position is most inaccurate in proximity to the pixel center $
|x_0| = {p\over 2} 
$.
In order to minimize that effect the sub-pixel division should comprise that higher uncertainty central region in a single sub-pixel.
An exemplary division to 3 sub-pixels in $x$ direction is shown in Fig.~\ref{fig:Difference_ratio_cumsum}.

In conclusion, the assignment of sub-pixel coordinates to a given photon event consist only of two steps: (i) calculation of $d_x$ and $d_y$ with formulas~\eqref{eq:d_x} and~\eqref{eq:d_y}, and (ii) comparison of obtained values with sub-pixel borders expressed in pixel ratio coordinates. 
As a result the algorithm 
is fast and can be applied on-line.

\section{Experimental \label{sec:Experimental}}

\begin{figure}[!b]
\centering
\includegraphics[width=\linewidth]{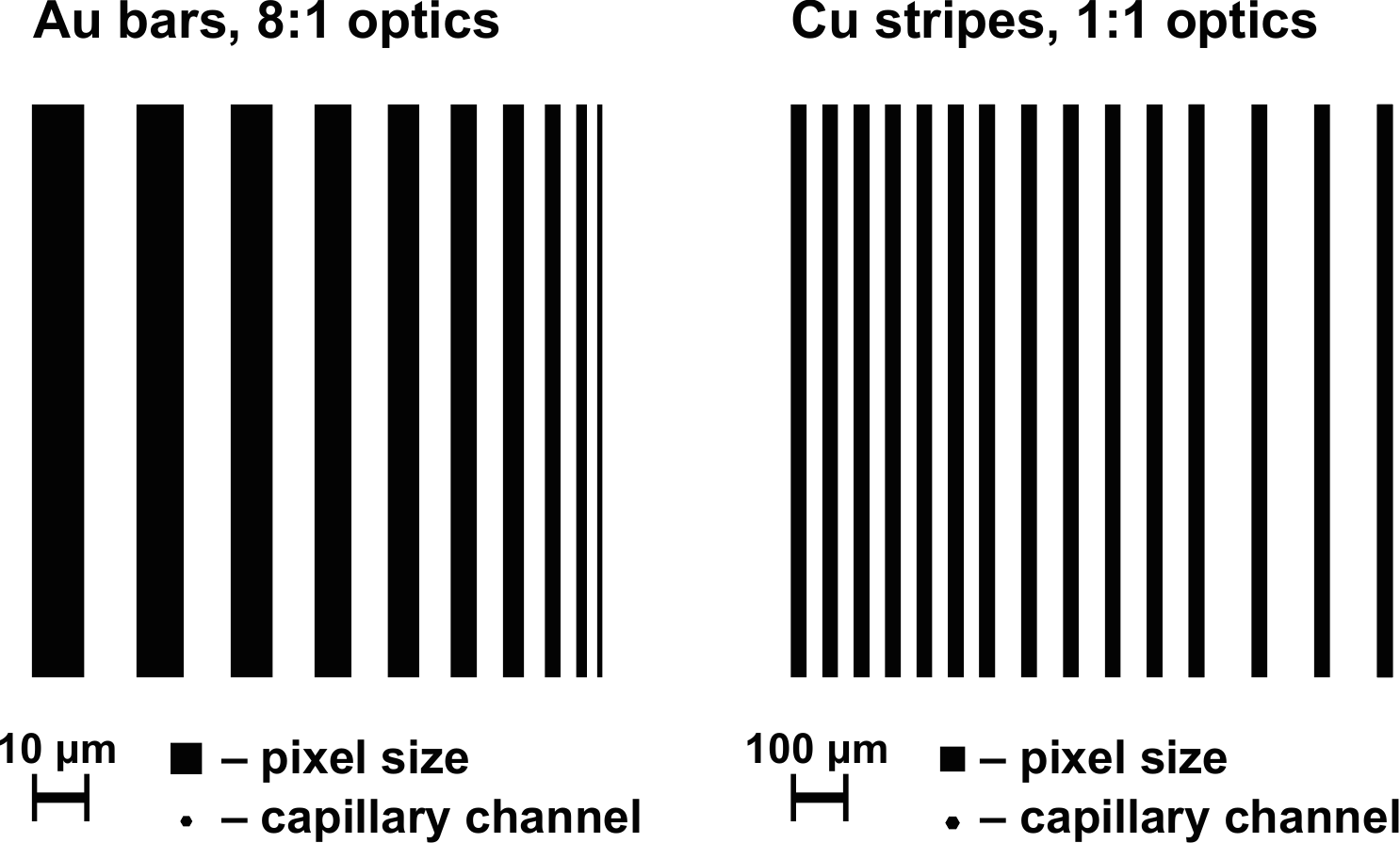}
\caption{
Sketch of 
Au bar pattern (left) and Cu stripes (right).  
Structures were measured with, respectively,  focusing 8:1 and  parallel 1:1 optics;
sizes of corresponding pixels (side length of, respectively, \unit{6}{\mu m} and \unit{48}{\mu m}) and capillary channels (entrance diameter of, respectively,  \unit{2}{\mu m} and \unit{24}{\mu m}) are shown for comparison.
%
 }
\label{fig:structures}
\end{figure}

To examine the performance of the sub-pixel algorithm two test samples were measured.
The first one  contains several microns of Au on a Si support arranged in bar-like patterns and a uniform reference layer.\cite{Strub2008}
The bar pattern consists of 10 Au lines with the width and the spacing between them decreasing from 10 $\mu$m to 1 $\mu$m in a 1 $\mu$m steps.
The resolution limit  can be easily found as the width of the narrowest 	recognizable line.
The second structure was produced by the Fraunhofer-Institut f\"ur Zuverl\"assigkeit und Mikrointegration (IZM) in Berlin and consists of \unit{3}{\mu m} thick and \unit{30}{\mu m} wide  Cu stripes deposited on Si wafer with 200~nm TiW adhesion layer.
The stripes are aligned in parallel and distributed  in groups with constant spacing of 30, 50 and \unit{90}{\mu m}.
Sketches of both structures are illustrated  in  Fig.~\ref{fig:structures}.

Imaging was performed with two types of polycapillary lenses -- a high-resolution conical  optics with a 8:1 magnification  that was employed for the Au pattern, and a parallel 1:1 optics used for the Cu stripes structure. 
The magnification optics has  a single channel entrance diameter of $d_{in} = \unit{2}{\mu m}$.
The exit diameter of a channel span to $d_{out} = \unit{16}{\mu m}$ which is  one third  of the pixel size ($p=\unit{48}{\mu m}$).
A single channel diameter of the 1:1 optics amounts to $d = \unit{24}{\mu m}$ which is   half of the pixel size.

In order to ensure high enough photon count rate, measurements were performed  
at
synchrotron radiation facility with the beam
provided by the BAM{\it line} at BESSY~II, \cite{Riesemeier2005}
and at newly developed PIXE beam line (HS-PIXE) at Ion Beam Center
at Helmholtz-Zentrum Dresden-Rossendorf (HZDR).\cite{Nowak2014_Examples_of_XRF_and_PIXE_imaging}
Detailed descriptions of two setups can be found in Ref.~\citenum{Scharf2011} and~\citenum{Nowak2014_Examples_of_XRF_and_PIXE_imaging}.

For  imaging purpose  Au L and Cu K lines intensity distributions were assessed.
Valid photon events were selected with an algorithm described elsewhere. \cite{Scharf2011}
The algorithm rejects the noise such as cosmic rays and
accepts only photon events having an appropriate arrangement of pixels above the noise threshold. 

The position of the photon hit is first estimated by the weighted
position of the pixel with the highest intensity and its nearest neighbors.
Subsequently a  2$\times$2 pixel box  around a pixel corner closest to the estimated position
 is selected and the sub-pixel algorithm is applied. 
The probability functions  of pixel intensity ratios  in $x$ and $y$ directions were obtained directly from the measurements. 
Probability density  and cumulative probability functions of the intensity ratios in $x$ direction  for Au L photons are presented in Fig.~\ref{fig:Difference_ratio_histogram} and~\ref{fig:Difference_ratio_cumsum}.

\section{Results}

		\begin{figure*}
		\centering
		\includegraphics[width=.33\linewidth,trim = 2.7cm 0 2.0cm 0,clip ]{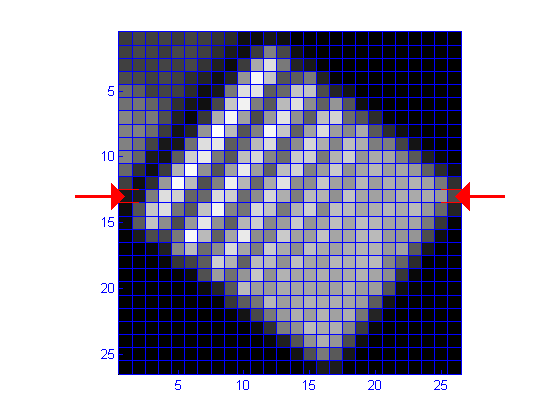}%
		\includegraphics[width=.33\linewidth,trim = 2.7cm 0 2.0cm 0,clip ]{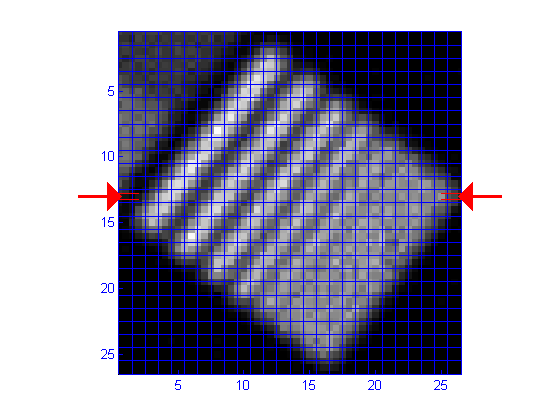}%
		\includegraphics[width=.33\linewidth,trim = 2.7cm 0 2.0cm 0,clip ]{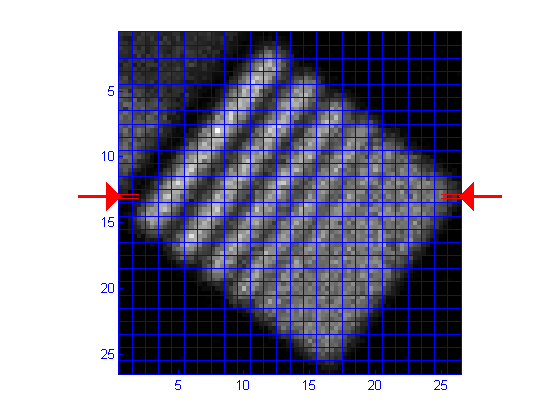}
		\caption{Normal   resolution (left), 2$\times$2 (center) and 3$\times$3 (right) sub-pixel division  images of the Au bar pattern.  
		The image was acquired at the BAM{\it line} at BESSY~II  with an 8:1 magnification lens. 
		 The grid shows the positions of the real pixels, each
	 corresponding to 6$\times$6~$\mu$m$^2$ area on the sample surface.
		 Red flashes indicate the (sub-)pixel row chosen for comparison in Fig.~\ref{fig:Au_stripes_pixel_profile}. }
		\label{fig:Au_stripes}
		\end{figure*}

	 \begin{figure}[p]
	   \centering
	   \includegraphics[width=1\linewidth]{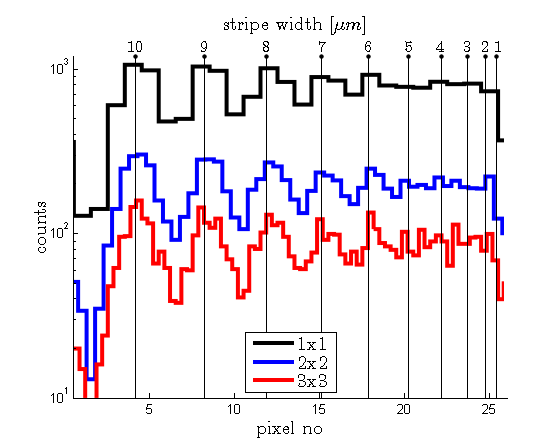}
	   \caption{Pixel intensity profiles  along  one sub-pixel row indicated in Fig.~\ref{fig:Au_stripes} with  red rims.  
	   The profiles are plotted for normal resolution (black line), 2$\times$2 (blue line), and 3$\times$3 (red line) sub-pixel division.
	   	Estimated positions of the Au bar centers are indicated with black vertical lines. 
	   On top the actual Au stripe widths are given.  
	     }
	   \label{fig:Au_stripes_pixel_profile}
	   \end{figure} 

	\begin{figure}[p]
	\centering
	\includegraphics[width=1\linewidth]{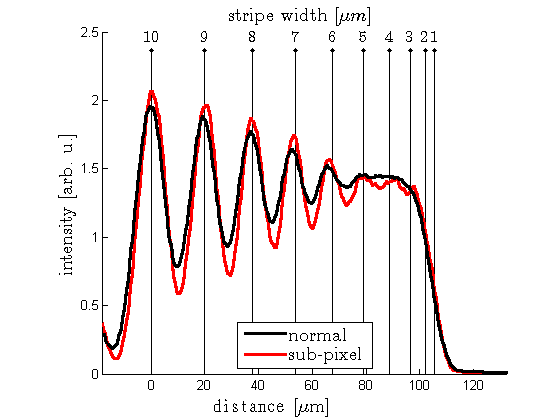}
	\caption{Intensity profile across Au bar structures  calculated for normal resolution (black line) and 3$\times$3 sub-pixel division (red line) images. 
	Estimated positions of the Au bar centers are indicated with black vertical lines. 
	The intensity profile was constructed from data sampled at regular intervals across the Au bars. 
	To increase the signal to noise ratio the intensity was integrated over the whole length of Au bars. 
	}
	\label{fig:Au_stripes_profile}
	\end{figure}

\subsection{Au bar patter}
In Fig.~\ref{fig:Au_stripes} the image of Au bar pattern is presented in normal resolution and with pixels divided to 2$\times$2 and 3$\times$3 sub-pixels.
The normal resolution image exhibits strong pixelation introducing artificial jerking of the Au bars and decreasing the contrast between structure and background. 
Only the 4 thickest stripes can be distinguished in the original image; the rest is blurred.
With the division into 2$\times$2 sub-pixel another Au line is resolved.
The 3$\times$3 subpixel division, however, does not introduce any visible improvement.
In this regime a single sub-pixel correspond to \unit{2}{\mu m} distance on the sample which is exactly the entrance channel diameter of the polycapillary optics (\unit{2}{\mu m}).
In addition the small effective area of a sub-pixel significantly lowers the count rate and increases the noise contribution.

In Fig.~\ref{fig:Au_stripes_pixel_profile} three intensity profiles along a single \mbox{(sub-)}pixel row  spanning along the whole Au bar structure are depicted.  
A profile intensity drop with pixel division is clearly visible.
This is an effect of the sub-pixel area decrease. 
For a pixel divided to $n$$\times$$n$ sub-pixels the intensity is decreased $n^2$ times. 
However,  the increased sampling frequency   evidently increases the  contrast. 
For the profile with no  sub-pixel division only the 4 thickest Au bars can be definitely differentiated. The intensity valley between $6 \ \mu$m and $5 \ \mu$m bars is  not visible. 
In case of 2$\times$2 pixel division the \unit{6}{\mu m} bar can be clearly distinguished. 
For 3$\times$3 sub-pixel division, due to the considerable noise content, there is no significant  amendment.

It should be noted that due to the Nyquist-Shannon sampling theorem, \cite{Shannon1949}  details smaller than twice the sampling distance cannot be correctly  distinguished.
The theorem says that making $n$ samples
over a certain distance only a signal comprising
less than $n/2$ elements can be  rendered properly.
Higher number of elements will lead to signal aliasing
resulting in distortions and artifacts.
Of course, with the sub-pixel division, this constrain is relaxed. 

As a single pixel in a 8:1 magnified image represents a square with a side length of $6 \ \mu$m the camera in standard resolution  
cannot correctly represent features smaller than $12 \ \mu$m.
Still, as can be seen in the presented example, even without sub-pixel resolution much smaller lines  can be distinguished; though with some alterations.
For instance in Fig.~\ref{fig:Au_stripes_pixel_profile}  the normal resolution profile shows  the \unit{8}{\mu m } bar to be thicker than the \unit{9}{\mu m } one.

In Fig.~\ref{fig:Au_stripes_profile} two additional intensity profiles are shown.
The profiles represent the cumulative intensity across the lines parallel to Au bars sampled at regular intervals.
The sampling interval was identical for both curves corresponding to normal resolution and 3$\times$3 sub-pixel division;
therefore the relative intensity of two profiles does not differ.
For the sake of legibility the profile corresponding to 2$\times$2 sub-pixel division is not present.

The tilt of the Au bar structure with respect to the pixel lines eliminates the effects due to pixelation.
As a consequence the \unit{6}{\mu m} Au bar can be  resolved even in case of normal resolution.
Also here the  contrast improves when sub-pixel resolution is applied.
In addition the enlarged integration area increases the statistics leveling the noise contribution.
As a result the sub-pixel resolution profile allows distinction of  \unit{5}{\mu m} or even \unit{4}{\mu m}  bars.

   \begin{figure}
   \centering
   \includegraphics[width=1\linewidth]{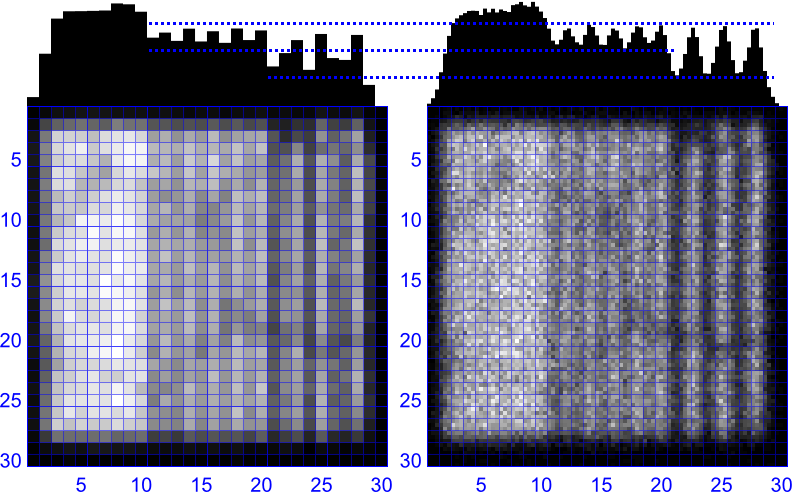}
   \caption{
   Normal resolution (left) and 3$\times$3 sub-pixel resolution (right) images of the Cu stripes pattern. 
   The image was acquired at HS-PIXE beam line at HZDR with a parallel 1:1  optics. 
   The grid shows the positions of the real pixels, each
   corresponding to 48$\times$48~$\mu$m$^2$ area on the sample surface.
   On top  the cumulative intensity along each (sub-)pixel column  is presented. 
   The dotted guide lines indicate the contrast levels  for  the sub-pixel resolution image.
     }
   \label{fig:Cu_stripes}
   \end{figure}

\subsection{Cu stripes}
The consequences of too sparse sampling are clearly visible in the image of Cu stripes (see Fig.~\ref{fig:Cu_stripes}).
The image represents a structure with three groups of Cu stripes with  dimensions below the Nyquist-Shannon resolution limit. 
Here the  structure is  aligned with the pixels and the effect of aliasing is clearly visible. 
Normal resolution and 3$\times$3 sub-pixel division images are presented for comparison.

The left most group is too confined and individual stripes cannot be distinguished even when  sub-pixel resolution is applied.
The structure has stripes with dimensions and spacing of \unit{30}{\mu m} which is too close to the single capillary  channel diameter (\unit{24}{\mu m}).

The stripes in the central group are distributed with  interval of \unit{90}{\mu m}, which is  comparable to the Nyquist–Shannon resolution limit, {\it i.e.}, double the pixel size ($p=\unit{48}{\mu m}$).
In the standard image the stripes are merely distinguishable  and  confined to a single pixel line. 
The sub-pixel resolution image shows a much better representation of the structure.

Finally, the group on the right has an interval large enough to  fulfill the  Nyquist-Shannon theorem, but  individual stripes are still below the limit.
Therefore the structure is clearly resolved, however the stripe's  widths and positions are not rendered well giving a false impression of a patchy structure.
The sub-pixel resolution image correctly presents uniform, equidistant stripes.

For both central and right most group of stripes there is a noticeable increase of the contrast level when sub-pixel resolution is used.
However, the contrast of the central group is much smaller. 
This is not surprising as the dimensions in the central group are much closer to the size of a single polycapillary channel which is an ultimate resolution limit.

 \subsection{Au layer}
 
  	\begin{figure*}
  	\centering
  	\includegraphics[width=\linewidth ]{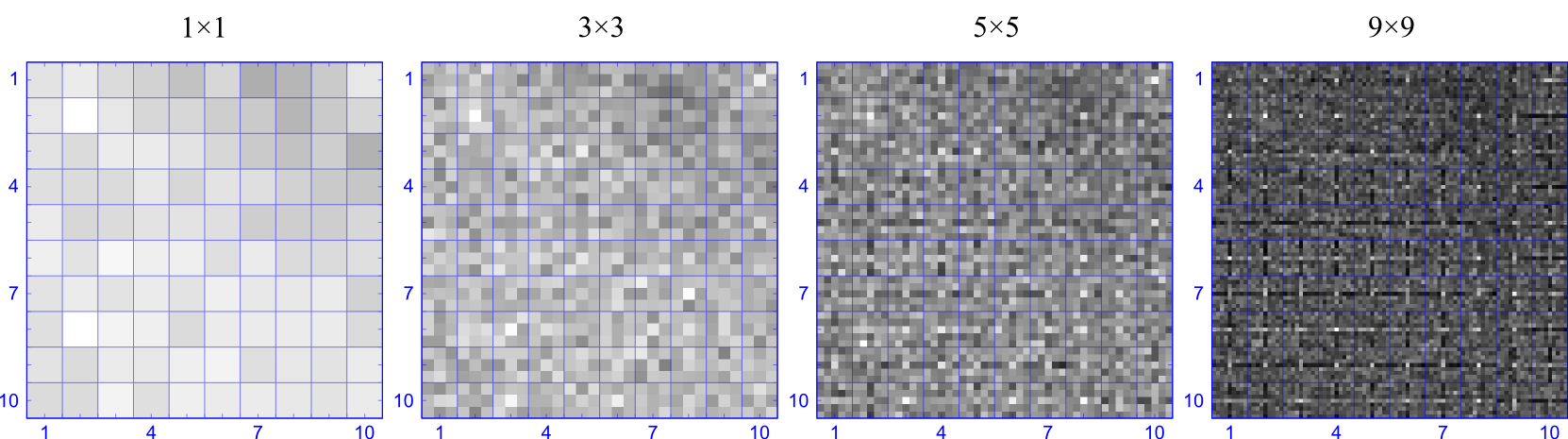}%
  	\caption{Normal   resolution and 3$\times$3,  5$\times$5, and 9$\times$9 sub-pixel division  images of a Au layer. 
  		The image was acquired at the BAM{\it line} at BESSY~II  with a 8:1 magnification lens. 
  	Artificial intensity structure following the pixel center position is strongly indicated for the sub-pixel resolution image.    
  	 The grid shows the real pixel positions.
  	 }
  	\label{fig:artefacts}
  	\end{figure*}
 
In order to test the limits of sub-pixel resolution algorithm an additional assay was performed on a reference Au layer. 
A small portion of the image was compared for the case of normal pixel resolution and increasing sub-pixel division (see Fig.~\ref{fig:artefacts}).
As can be seen  very dense divisions lead to creation of intensity artifacts following the pixel center positions. 

This effect was already discussed in Section~\ref{sec:sub_pixel_division} and is a result of the uncertainty of a position of a photon hit reaching the middle area of a pixel.
Nevertheless, pixel divisions up to 5$\times$5 sub-pixels are usually not so much affected  by this  inconvenience.  
It should be noted that due to a nonzero divergence of  transmitted photons and a nonzero 
optics-detector distance in \SLcam  the footprint of a single polycapillary channel on the detector cannot decrease below  a dozen of microns. 
Thus, keeping in mind that the dimension of a pnCCD pixel  is \unit{48}{\mu m}, a pixel division into 5$\times$5 sub-pixels should be sufficient to reach the optimal limits of resolution.

\section{Summary}

We propose a modified approach to sub-pixel resolution algorithm taking into account all the photon events occurring on a CCD plane. 
The sub-pixel position of the photon hit is assessed from the pixel footprint of the generated electron cloud. 
The calculations are performed based on the pixel intensity ratios of a  2$\times$2 pixel box holding  photo electrons. 
The method is independent from the actual shape and creation process of the charge cloud
and does not reject the photon events from the pixel center. 
The algorithm 
is fast and can be employed on-line.


%
%
%

The sub-pixel resolution was applied to several test structures. 
A notable enhancement in quality of the acquired images comprising contrast and resolution improvement, as well as  elimination of aliasing due to pixelation was demonstrated.
For images acquired with  8:1 magnifying  optics
a resolution limit of \unit{5}{\mu m} was assessed.
Due to an electronic noise and random variations of electron cloud shape the sub-pixel coordinates cannot be given with unlimited precision. 
It was shown that in case of   48$\times$\unit{48}{\mu m^2} pixel pnCCD the  5$\times$5 sub-pixel division is a realistic limit of the method. 

\section*{Acknowledgments}

This work has been supported by Marie Curie Actions - Initial Training Networks (ITN) as an Integrating Activity Supporting Postgraduate Research with Internships in Industry and Training Excellence (SPRITE) under EC contract no. 317169.


%

\bibliographystyle{rsc}
\bibliography{bibliography}
\inputencoding{utf8}

\end{document}